\documentclass[a4paper,10pt,twocolumn]{article}
\usepackage[utf8]{inputenc}
\usepackage[T1]{fontenc}
\usepackage{mathptmx}

\usepackage{amssymb,amsmath}
\usepackage[english]{babel}
\usepackage{indentfirst}
\usepackage{subfig}
\usepackage{abstract}
\usepackage{authblk}
\usepackage{verbatim}

\usepackage{graphicx}

\begin{document}

\bibliographystyle{plain}

\normalfont

\twocolumn[
\title{\large \bfseries Structural and dynamical properties of nanoconfined supercooled water}
\author{Oriol Vilanova, Giancarlo Franzese \and \itshape \normalsize Departament de Física Fonamental, Universitat de Barcelona, \itshape Av. Diagonal 647, 08028 Barcelona, Spain}
\date{}
\maketitle
\begin{onecolabstract} 
Bulk water presents a large number of crystalline and amorphous ices.
Hydrophobic nanoconfinement is known to affect the tendency of water
to form ice and to reduce the melting temperature. However, a systematic 
study of the ice phases in nanoconfinement is hampered by the computational
cost of simulations at very low temperatures. Here we develop a
coarse-grained model for a water monolayer in hydrophobic nanoconfinement  
and study the formation of ice by Mote Carlo simulations. We find two ice 
phases: low-density-crystal ice at low pressure and high-density hexatic 
ice at high pressure, an intermediate phase between liquid and 
high-density-crystal ice.\\
\end{onecolabstract}
]

\section{Introduction}

Water phase diagram is very complex when compared to other 
liquids. For example, water is a polymorph with an unusually large number of solid
phases (crystalline and amorphous ices): more than 20 with the last phase 
discoverered in 2009 \cite{IceXV}. 
Formation of ice in confined systems is a relevant
subject in nanocience and biology, in areas like cryopreservation of
food and human tissues or cells. Due to the property of water to
expand when the liquid transforms into ice, the formation of ice in
confinement can drammatically damage or destroy the confining
structure. Therefore, it is important to understand the properties of
ice in nanoconfinement, especially for 
hydrated systems at low temperature where water is highly confined, as
for example in biological cells or on the surface of proteins at low hydration level
\cite{FoodBio,TwoCrossovers}. 

Simulations can help to answer open questions in this fields, but they
are hampered by the large computational costs of calculations for large
systems at low temperature with detailed models of water. With the aim
of developing a model that preserve essential properties
of water and is also computationally efficient, here we perform Monte
Carlo simulations of a coarse-grained model of water. In its original
formulation the model allows for very
efficient simulation studies. On the other hand, it has a simple hamiltonian 
that allows for theoretical studies
\cite{TheoryDensity,HBCoop}, but does not allow the study of structural
properties because the positions of water molecules are
coarse-grained. Here, we extend the model, introducing the coordinates of
the molecules $\vec{r}_{i}$ to perform the structural analysis. 
We study the phase diagram and, in particular, how the hydrophobic
nanoconfinement affects the ice formation for a water monolayer.
We find two forms of ice and we characterize their structure.

This work is organized as follows: we introduce the Model in 
the first section and give details about the Monte Carlo method in the
second section; we present the results in the third section,
discussing the structural and dynamical properties; we make our
conclusions in the final section.


\subsection{The Model}

We consider a water monolayer confined 
between two hydrophobic plates at separation $h\simeq 1nm$.
The hydrophobic interaction with the plates is
schematically represented as purely repulsive. 
By molecular dynamics simulations of a detailed model of water, it has been 
shown that a water monloayer under these conditions forms a
two-dimensional ice with square symmetry
\cite{Thermodynamics,Monolayer,BilayerIce}. Therefore, we adopt a square
partition to coarse grain the 
confined water, divinding the total occupied volume $V$ into $N$ cells of square
section and height $h$. Each cell has a volume $v=V/N$ and the coarse
grain is made with the hypothesis that the system is homogenous and
 each cell  contains one single
water molecule.  

We consider the case in which pressure $P$,
temperature $T$ and number of molecules $N$ are fixed, and  the total
volume $V$ can vary. Therefore, the volume per molecule $v$, and the
number density $\rho\equiv 1/v$, are functions of $P$ and $T$ at any
fixed $N$.  


The hamiltonian of the model has several water-water interaction terms. 
The first is the isotropic Van der Waals interaction, due to
dispersive attractive forces and short-range repulsive interactions,
represented by a Lennard-Jones potential:
\begin{equation}
\mathcal{H}_{\text{VW}} \equiv \sum_{i < j} \varepsilon 
\left[
  \left(\frac{r_{0}}{r_{ij}}\right)^{12} -
  \left(\frac{r_{0}}{r_{ij}}\right)^{6} 
\right]
\end{equation}
Here $\varepsilon\equiv 5.8$~kJ/mol is the characteristic energy,
$r_{0}\equiv 2.9$~\AA\ is the diameter  
of the molecules and $r_{ij}$  the distance between two 
water molecules in the cells $i$ and $j$.
In order to reduce the computational cost of the simulations,
we introduce a cutoff distance at $r_{\rm cutoff} = 2.5 r_{0}$ and add a linear
 term that set to zero the potential at $r_{\rm  cutoff}$.

Two neighbouring molecules can form a hydrogen bond when the OH---O
distance is less than \mbox{$r_{\rm max}-r_{\rm  OH}=3.14$\AA}, and if
${\widehat{\rm OOH}}<30^{\rm o}$. To account for this interaction the model
includes a term
\begin{equation}
\mathcal{H}_{\text{HB}} \equiv
-J N_{\text{HB}}
\end{equation}
where $J\equiv 2.9$~kJ/mol and
$N_{\text{HB}}\equiv \sum_{\langle i,j  \rangle} \beta_{ij}$ 
with \mbox{$\beta_{ij} \equiv \delta_{\sigma_{ij},\sigma_{ji}} \Theta(r_{ij} - r_{\rm max})$},
and $\Theta(x)\equiv 1$ if $x>0$, or $0$ otherwise.
 Each molecules $i$ has a  bond indices \mbox{$\sigma_{ij}
\in \{0,1,\dots q-1\}$}  for each nearest neighbor molecule $j$. The
choice $q = 180º/30º = 6$ accounts correctly for the entropy loss
associated with the formation of a hydrogen bond because by definition 
$\delta_{\sigma_{ij},\sigma_{ji}}\equiv 1$ 
if $\sigma_{ij}=\sigma_{ji}$, $\delta_{\sigma_{ij},\sigma_{ji}}\equiv
0$ otherwise. 
The notation  $\langle i,j\rangle$ 
denotes that the sum is performed over nearest neighbors, implying
that each molecule cannot form more than 4 bonds with its nearest neighbours. 

When water molecules form a hydrogen bond network, the resulting
configuration occupies more space than at close packing.
This effect is included in the model as a volume increase per formed bond
equal to
$v_{\text{HB}}=0.5v_0$, correspondig to the average density increase
between high density ices VI and VIII and low density (tetrahedral)
ice Ih in bulk water \cite{Soper,WaterStructure}.
The total volume of the system is, therefore,
$V = V_0 + v_{\text{HB}} N_{\text{HB}}$, where $V_0$ is the volume
when there are no hydrogen bonds.

As an effect of cooperativity, the O-O-O angle distribution becomes sharper 
at lower $T$, reducing the possible orientations of the molecules \cite{HBCoop,CooperativeDomains,IntramolecularCoupling}. 
This cooperative term, resulting from three-body interactions, is accounted
for by the term
\begin{equation}
\mathcal{H}_{\text{coop}} \equiv -J_{\sigma} \sum_{i} \sum_{(k,l)_{i}} \delta_{\sigma_{ik},\sigma_{il}}~,
\end{equation}
where $J_\sigma\equiv 0.29$ kJ/mol,  and $(k,\ell)_i$ indicates each of
the six different pairs of the four 
bond-indices $\sigma_{ij}$ of a molecule $i$.  
The effect of this term is to locally drive the molecules toward an ordered
configuration.

In its original formulation the model is defined by coarse-graining
the molecules coordinates $r_{ij}$ with the center of each cell. 
Furthermore, the effect of the cooperativity on the O-O-O angle
distribution  is taken into account in terms of the associated 
entrophy change, but not in terms of angular coordinates.
Therefore, no detailed structural analysis is possible. To allow the
calculation of the 
radial distribution function $g(r)$ and the angular distribution
function $g(\theta)$, in the following subsection we extend the
original model introducing a term that explicitly depends on these
variables. 

\subsubsection{Extension of the model}

The new Hamiltonian term is a three-body interaction that 
depends on the formation of hydrogen bonds between triads of molecules 
and their relative angles $\theta^{i}_{kl}$:
\begin{equation}
\mathcal{H}_{\theta} \equiv J_{\theta} \sum_{i} \sum_{\langle\langle
  k,l\rangle\rangle _{i}} \beta_{ik} \beta_{il} \Delta
(\theta_{kl}^{i}),
\end{equation}
where $J_{\theta} = 0.5\varepsilon$. 
The sum is over all the neighbouring pairs of molecules $k$ and $l$ that are 
bonded to the molecule $i$, with the restriction that $k$ and $l$ 
must be second nearest neighbors to each other.
The function $\Delta(\theta)$ is a smooth function 
of the angle between the centers of the three molecules with a minimum at 
$\pi/2$.
We adopt this choice because molecular dynamics simualtions of a
detailed water model show that, under the conditions considered here,
a confined water monolayer forms a square crystal \cite{Thermodynamics,Monolayer}. 
We chose 
\[ \Delta(\theta) \equiv \frac12  \left[  1 - \cos(4 \theta)
\right], \]
which is a non-negative function in $[0,1]$ with minima at $\pi/2$ and
$\pi$, and we approximate it with
\begin{equation}
\Delta(\theta) \simeq 4(\theta - \pi/2)^{2}
\end{equation}
around $\theta^{i}_{jk} \simeq \pi/2$.

This value of $J_{\theta}$ is set to avoid the formation of bonds when
the $\theta^{i}_{jk} \approx 60^\circ$, because 
\[\mathcal{H}_{\theta}(\theta^{i}_{jk} = \pi/3) =
\mathcal{H}_{\theta}(\theta^{i}_{jk} = 2\pi/3) \simeq 4 N J_{\theta},\]  
and
\[\mathcal{H}_{\theta} + \mathcal{H}_{\text{J}} \simeq 4 NJ_{\theta} + 2
N J = 2 N \varepsilon - N \varepsilon = N \varepsilon > 0.\]
Therefore,  the formation of hydrogen bonds is energetically
unfavourable when $\theta^{i}_{jk} \approx 60^\circ$.

The total hamiltonian of the model is
\begin{equation}
\mathcal{H} \equiv \mathcal{H}_{\text{VW}} + \mathcal{H}_{\text{HB}} + \mathcal{H}_{\text{IM}} + \mathcal{H}_{\theta}.
\end{equation}

\subsection{Metropolis MC method}

We perform MC simulations at constant number of 
molecules  $N= 900$ and fixed pressure $P$ and temperature $T$, 
allowing fluctuations of the volume $V$. 
One MC step consists in updating $5N + 1$ variables: $N$  
vectors $\vec{r}_{i}$ describing the position of molecules with
respect to the center of their cell, $4N$ bondig indices
$\sigma_{ij}$ and the total volume $V_0$. 
We adopt the Metropolis algorithm: we choose one of the 
$5N+ 1$ variables at random and attempt to change its state  to a new random value.
We accept the new state with
probability $\exp [-\beta \Delta G]$ if $\Delta G > 0$ 
and with probability 1 otherwise. Here $\beta\equiv 1/k_BT$, $k_B$ is
the Boltzamann constant, $\Delta G\equiv G_{\rm new}-G_{\rm
  old}$ is the change in Gibbs free energy if the new state is
accepted, and 
\begin{align}
G &\equiv U + PV - TS \nonumber \\
&= \mathcal{H}+ PV - Nk_{\text{B}} T \log (V).
\end{align}

For a bonding index $\sigma_{ij}$, the new state is choosen at random among the $q=6$
possible states.
For each components of $\vec{r}_{ij}$, the new value is set to
$r_{\alpha,\text{new}} = r_{\alpha,\text{old}} + \epsilon_r$, 
where $\epsilon_r \in [-\delta r,+\delta r]$ is a random number and
$\alpha=x$, $y$ (we do not change the component $z$ and consider it as
a coarse-grained variable).
The volume is updated with a random change $V_0^{\text{new}} =
V_0^{\text{old}} + \epsilon_{V}$  
where $\epsilon_{V} \in [-\delta V, + \delta V]$ is a random number
\cite{NpTMC}.  
The parameters $\delta r$ and $\delta V$ are adapted in such a way to 
keep the acceptance ratio $\approx 40\%$ (adaptive step size algorithm)
\cite{OptimumMC,EfficientMC}. 

At any $P$, we equilibrate the system from random configurations at high
$T$ for $10^5  \sim 10^6$ MC steps and calculate the
thermodynamic averages over the following $10^6 \sim 10^7$ MC steps.
Keeping $P$ constant, we perform an annealing, i.e. we decrease the 
temperature $T$ 
a few K and, starting  
from the last configuration at the previous temperature, we use the
same statistics for equilibration and calculation of the averages.
To take into account the correlation of the data for the calculation
of the error on the estimates, we perform blocking averages where the 
size of each block depends on $P$ and $T$ and  is determined as twice
the number $\tau$ of MC steps needed to have uncorrelated data. The
number $\tau$ is estimated from the autocorrelation functions
introduced in Section \ref{corr}.

\section{Results}

\subsection{Phase Diagram}

\begin{figure}
\centering
\subfloat[]{\includegraphics[scale=0.6]{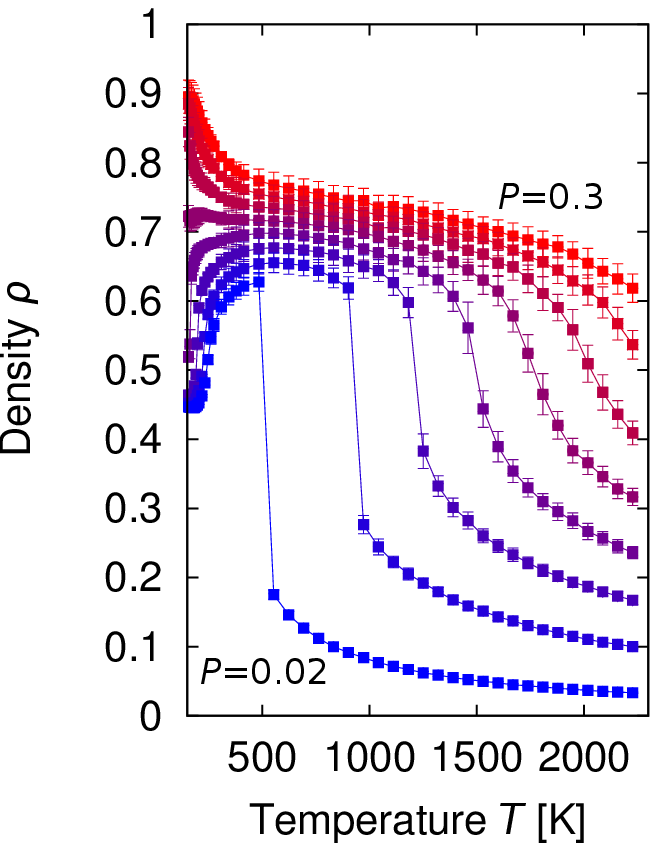}}\ \
\subfloat[]{\includegraphics[scale=0.6]{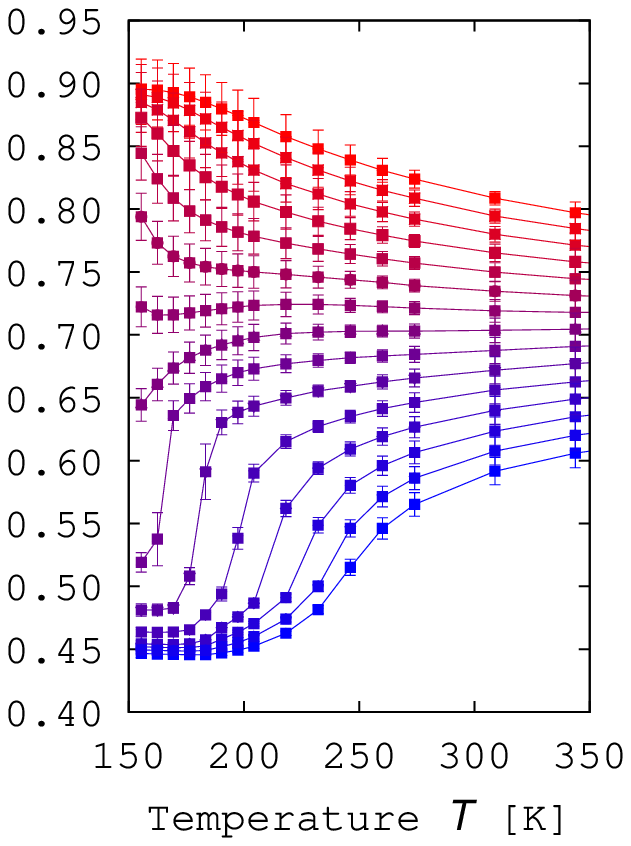}}\\
\subfloat[]{\includegraphics[scale=0.6]{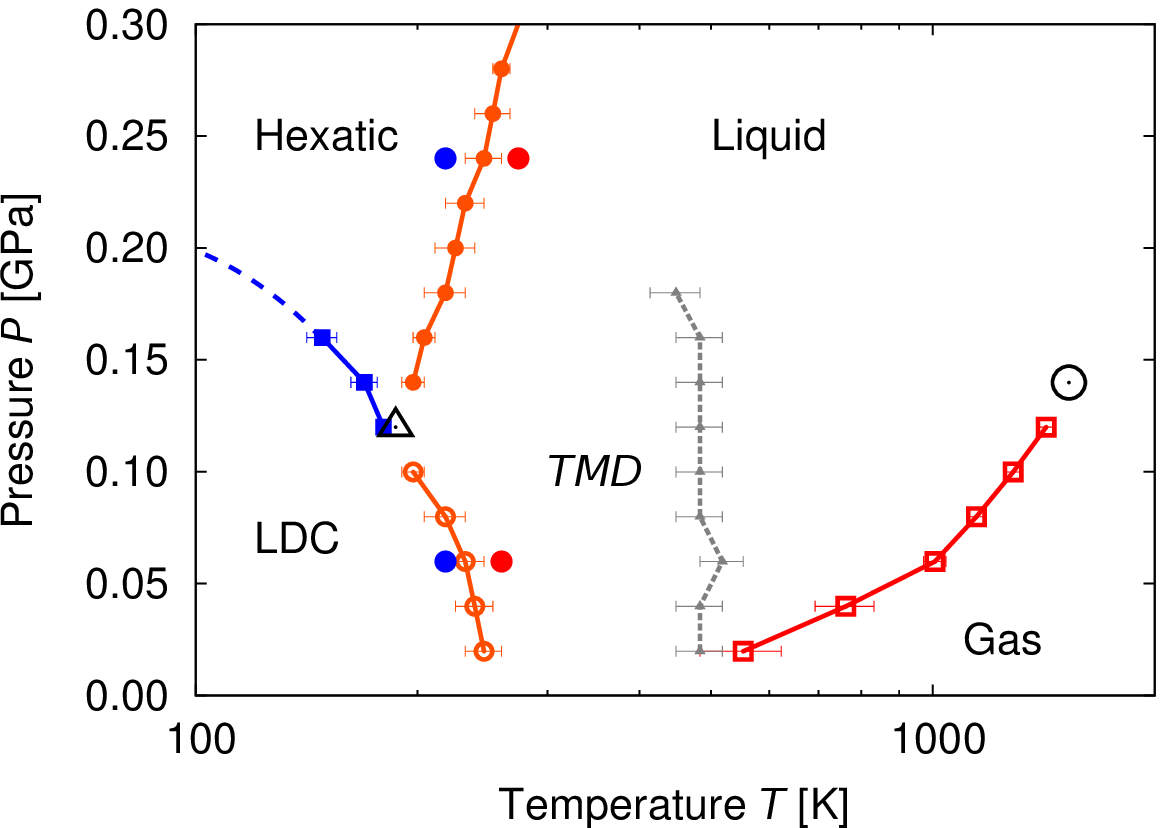}}
\caption{(a) Isobars for the confined water monolayer
in the $\rho-T$ phase diagram display a  discontinuity that marks  the
  first-order liquid-gas phase transition ending in a critical point 
(at $P=0.16\pm 0.02$~GPa, $T=1500\pm 200$~K)  and a density maximum for
$P<0.20$~GPa. Isobars are for  $P\in [0.02, 0.3]$~GPa separated by $0.04$ GPa 
(from the bottom). 
(b) At low $T$ a sudden decrease in $\rho$ suggests the occurrence of
another phase transition. As we discuss in the analysis, we associate
this sudden change in $\rho$ to the first-order phase
transition between the liquid and a low density crystal (LDC). Pressures 
are as in (a) but with a separation of $0.02$ GPa. (c) The $P-T$ phase 
diagram with gas-liquid coexistence line ending at the gas-liquid critical 
point (open circle) and the line of temperatures of maximum
density (TMD).
At low $T$ and low $P$, the liquid--LDC coexistence line merges, at
about $P=0.11\pm 0.01$~GPa and $T=190\pm 20$~K (open triangle), with a
line with positive slope corresponding to the liquid-hexatic phase
transition, as described in the text. 
Full circles are the state points of the distribution functions in Fig.s \ref{RDF},\ref{angular}.}
\label{phase-diagram}
\end{figure}

The phase diagram of the model displays the liquid-gas first-order phase 
transition ending in a critical point  (Fig.\ref{phase-diagram}). 
In the liquid phase we observe that at any $P<0.2$~GPa the density is
non monotonic. The locus of temperatures of maximum 
density (TMD) follows a line in the $P$-$T$ phase diagram, reproducing
one of the characteristc anomalies of water.

At low $P$ and low $T$ we find a rapid decrease of density $\rho$. This is the
consequence of  formation of a large number of 
hydrogen bonds and the cooperative reorientation of the 
molecules into a crystal configuration. In the next section we
caracterize this crystal as low-density crystal (LDC) with a  square
cell in its 2D projection.

At high $P$ and low $T$ our structural and dynamical analysis, presented in the next
sections, show that the system ``freezes'' into a solid. However, the
solid has no long-range translational order, but short-range
translational order and quasi-long-range orientational order. This is,
by definition, an ``hexatic'' phase, described by the theory of Kosterlitz,
Thouless, Halperin, Nelson, and Young \cite{Hexatic,MeltingLJ} for crystallization in
2D systems. The theory tells us that 
the hexatic phase is intermediate between the crystal and
the liquid phases and is separated by continuous phases transitions
with both phases. This is consistent with the fact that  we do not
observe any discontinuity in the density at high $P$ and low $T$
(Fig. \ref{phase-diagram}) and that our system is essentially in 2D
because we coarse-grain the $z$-component of the molecules. The
hexatic-liquid coexistence is 
characterized by the unbinding of disclinations, i.e.  lines of defects at which
rotational symmetry is violated.
The crystal-hexatic coexistence is where occurs the unbinding of
dislocations, i.e. particle-like topological defects.
The associated crystal phase is characterized by the same
orientational order of the hexatic phase, that is, as described in the
next section, hexagonal (or close-packing) and has a higher density of
the LDC. We therefore call it high-density crystal (HDC).
Finally, the structural analysis allows us to estimate the coexistence
line  between the LDC and the hexatic phase and  the triple point
where liquid, hexatic and LDC phases coexist (Fig. \ref{phase-diagram}).


It is interesting to observe that the phase diagram of our confined
monolayer reproduces  the change of slope of the ``ice'' line observed
for bulk water. The slope is negative at low $P$ and is positive at
high $P$. The ice phase at low $P$ is LDC characterized by
hydrogen bonds at $90^\circ$. At high $P$ the solid phase is hexatic,
where the number of hydrogen bonds is largely reduced and the water
interaction is dominated by the Lennard-Jones potential, as  in simple liquids.


\subsection{Structural Properties}

\subsubsection*{The radial distribution function}

\begin{figure}
\centering
\subfloat[Low pressure at $P=0.06$ GPa]{\includegraphics[scale=0.65]{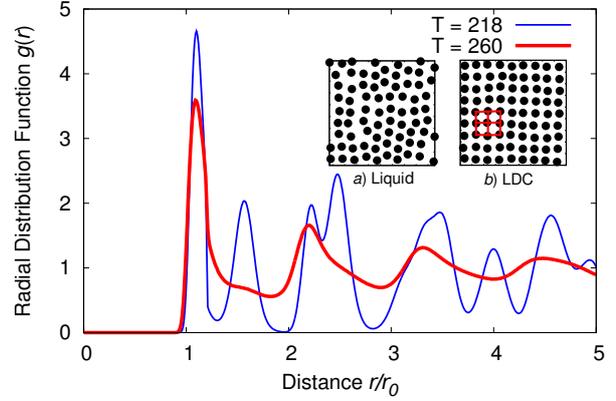}}\\
\subfloat[High pressure at $P=0.24$ GPa]{\includegraphics[scale=0.65]{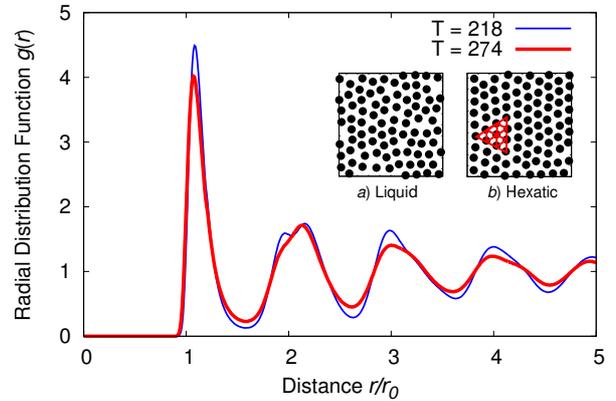}}
\caption{Radial Distribution Functions for the state points marked as 
coloured full circles in Fig.\ref{phase-diagram}. (a) At \mbox{$P=0.06$~GPa}
and $T=246$~K (in the liquid phase) and $T=218$~K (in the LDC phase).
Here and in the next panel, insets show a portion of typical
configurations at the state points represented in the main panels. 
The $g(r)$ for the LDC has many peaks correspoding to the long-range
translational order of the square crystal, while the $g(r)$ for the liquid
near the coexistence shows precursors of the LDC structure. 
(b) At $P=0.24$~GPa
and $T=274$~K (in the liquid phase) and $T=218$~K (in the hexatic phase).
The liquid $g(r)$ has a shoulder in the second peak that splits into a
small peak in the hexatic phase. The hexatic phase has liquid-like short-range
translational order due to the presence of many
disclinations, but crystal-like long-range orientational order, emphasized by
links in the inset describing the hexatic phase.}
\label{RDF}
\end{figure}

To study the static properties of the system we calculate
the radial distribution function (RDF) as
\begin{equation}
g(r) \equiv \frac{1}{\rho^{2}V} \sum_{i \neq j} \delta(r - r_{ij}).
 \end{equation} 
The quantity $g(r)2\pi r dr$ is 
proportional to the probability of finding a molecule at a distance
$r$ from a central one.

By crossing the ice-liquid lines of Fig.\ref{phase-diagram}c we can
observe structural changes in the $g(r)$.
At low $P$ (Fig.\ref{RDF}a), we find a large change in $g(r)$ within a
narrow range of $T$, marking the occurrence of the liquid-LDC
first-order phase transition. The LDC is characyerized by square
long-range translational order.

At high $P$ (Fig.\ref{RDF}b), we observe a shoulder in the second peak of the
$g(r)$ for the liquid. This shoulder develops into a small peak at
lower $T$. This structural change has been characterized \cite{Freezing} 
as the liquid-hexatic second-order phase transition. This
interpretation is consistent with the analysis of the typical
configurations at the lower $T$, showing liquid-like short-range
translational order and crystal-like long-range orientational
(hexagonal) order. The hexatic phase is the precursor of the HDC
close-packing crystal. The solid-like properties of this phase are
confirmed by the analysis presented in the next section.

Finally, at low $T$ by increasing $P$ the structural analys allows us
to locate the coexistence between the LDC and the hexatic phase. The
transition is charcterized by a sharp change of $g(r)$ indicating a
first-order phase transition between the LDC and the hexatic phase.

\subsubsection*{The angular distribution}

\begin{figure}
\centering
\subfloat[Low pressure at $P=0.06$ GPa]{\includegraphics[scale=0.5]{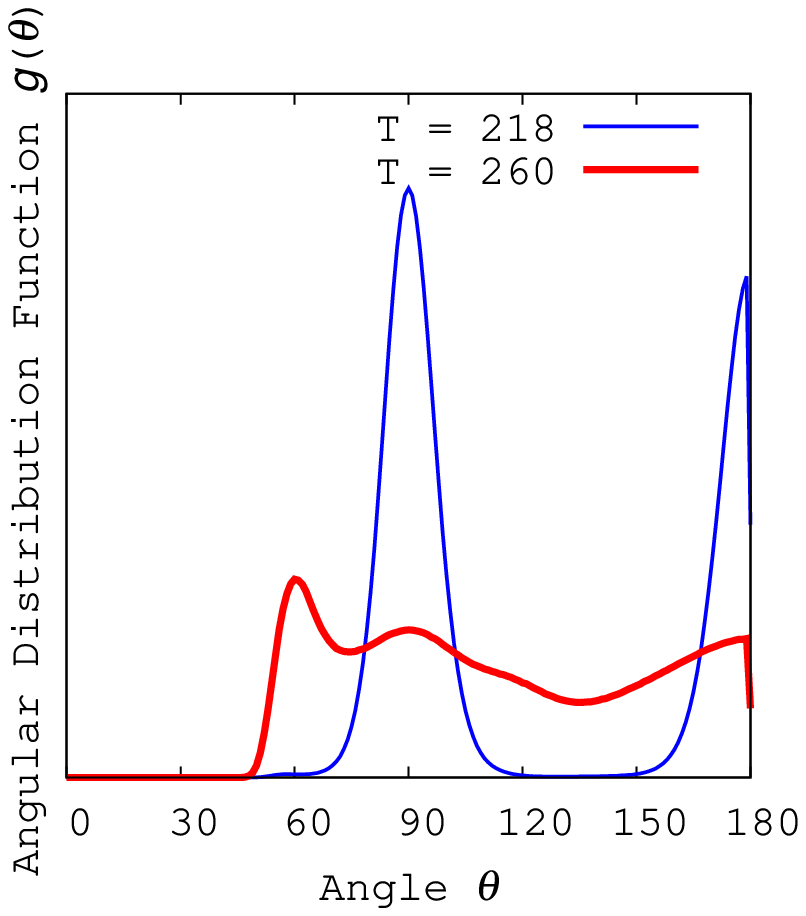}}\ \
\subfloat[High pressure at $P=0.24$ GPa]{\includegraphics[scale=0.5]{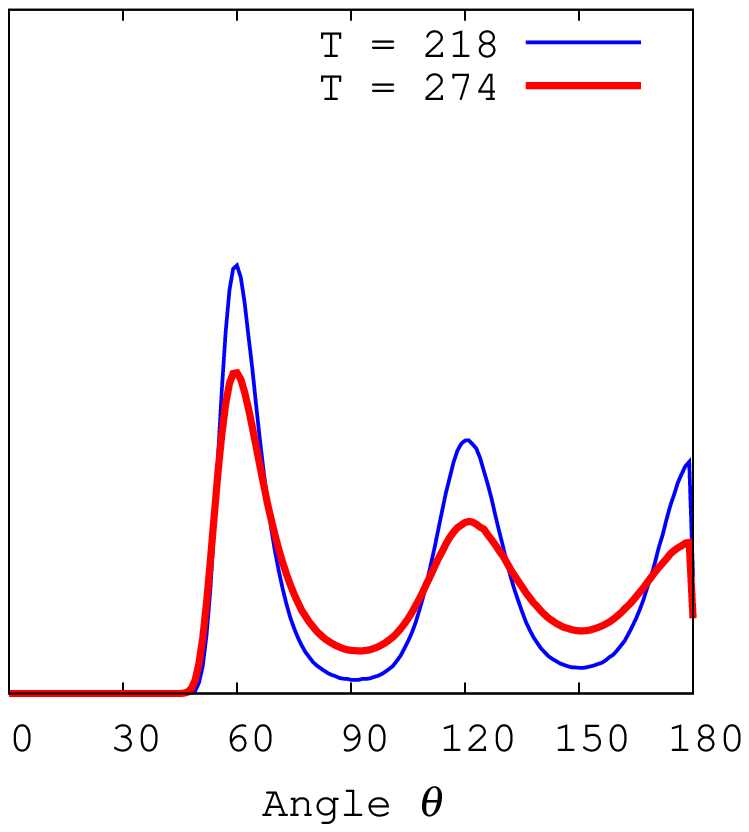}}
\caption{The angular distribution functions $g(\theta)$, calculated at the
  same state points as in Fig.\ref{RDF}, emphasizes the long-range
  orientational order in the LDC and the hexatic phases. 
}
\label{angular}
\end{figure}

To charcterize the structure, we also calculate the \mbox{O-O-O} 
 (not normalized)  angular distribution function, defined as
\begin{equation}
g(\theta) \equiv \frac{1}{N} \sum_{i=1}^{N} \sum_{(j,k)_{i}} \Theta(r_{ij} - r_{\rm max}) \Theta(r_{ik} - r_{\rm max}) \delta (\theta - \theta^{i}_{jk}).
\end{equation}
The quantity $g(\theta)r\ \rm{d}r\ \rm{d}\theta$ is
proportional to the probability  of finding two molecules 
($j$, $k$) at a distance $r \leq r_{\rm max}$ from a central one $i$ and 
forming an angle $\theta^{i}_{jk}=\theta$. The condition $r < r_{\rm
  max}$ limit our calculation to the first shell, in the condensed
phase, of the central molecule.

Our analysis of $g(\theta)$ (Fig.\ref{angular}) is complementary to
that of $g(r)$ and emphasizes the appearence of the long-range
  orientational order in the LDC and the hexatic phases. 
In the two solid phases,
the positions of the peaks are related to the 
symmetry of each crystal phase: the square LDC structure has peaks at 
$90^\circ$ and $180^\circ$, and the peaks of the hexatic solid phase,
with the same symmetry as the  HDC, are 
centered at $60^\circ$, $120^\circ$ and $180^\circ$.
In the liquid
  phase, instead, $g(\theta)$ never goes to zero showing the absence
  of orientational order.

\subsection{Dynamical Properties}

The study of the autocorrelation 
functions provide relevant informations about both the MC dynamics
and the transport properties of the system. From them we extract the
correlation times necessary to calculate in a correct way the
statistical errors of our observables. Moreover, from the correlation
time we estimate when the dynamics of the system is liquid-like or
solid-like. 


We calculate the hydrogen bonds autocorrelation function 
\begin{equation}
C_{M}(t) = \frac{1}{N} \sum_{i} \frac{\left< M_{i}(t_{0} + t)
    M_{i}(t_{0}) \right> - \left< M_{i} \right>^{2}}{\left< M_{i}
    ^{2}\right> - \left< M_{i}\right>^{2}}
\label{Cm}
\end{equation}
of the average molecular bonding index of molecule $i$
\mbox{$M_{i} \equiv \frac14 \sum_{j} \sigma_{ij} - (q - 1)/2$}. 
This quantity describes the hydrogen bonds dynamics of water
molecules. 

Next, we calculate the translational autocorrelation function
\begin{equation}
C_{r}(t) = \frac{1}{N} \sum_{i} \frac{\left< \delta\vec{r}_{i}(t_{0} + t) \cdot \delta\vec{r}_{i}(t_{0}) \right> - \left< \delta\vec{r}_{i} \right>^{2}}{\left< \delta r_{i} ^{2}\right> - \left< \delta\vec{r}_{i}\right>^{2}}
\label{Cr}
\end{equation}
where $\delta\vec{r} = \vec{r}_{i} - \vec{r}_{0,i}$ is the displacement of each molecule 
$i$ from the center of its cell.
In Eq.~(\ref{Cm}) and (\ref{Cr}) the time $t$ is measured in MC steps
and can be related to real time only by comparison with experiments. 
For example, it can be shown that the conversion factor between a MC
step and real time unit rescales  logarithmically
with $T$ at ambient pressure \cite{TwoCrossovers}.


\begin{figure}
\centering
\subfloat[HB autocorrelation]{\includegraphics[scale=0.4]{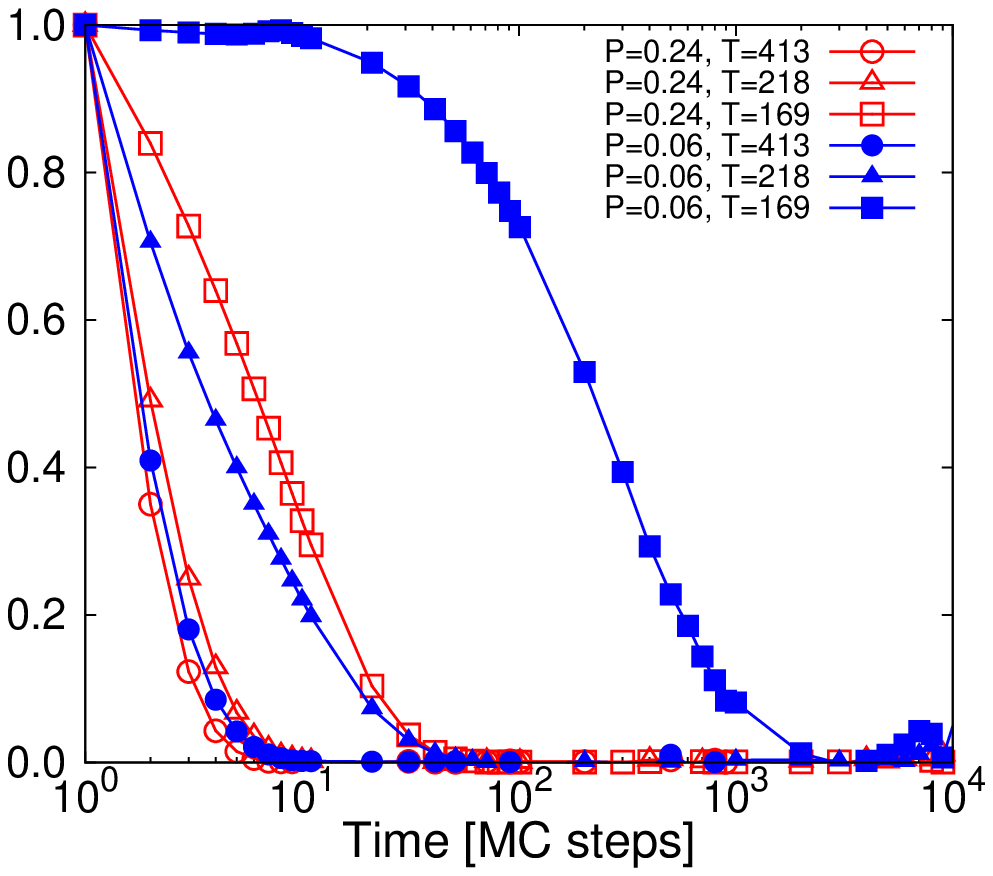}}\
\subfloat[Translational autocorrelation]{\includegraphics[scale=0.4]{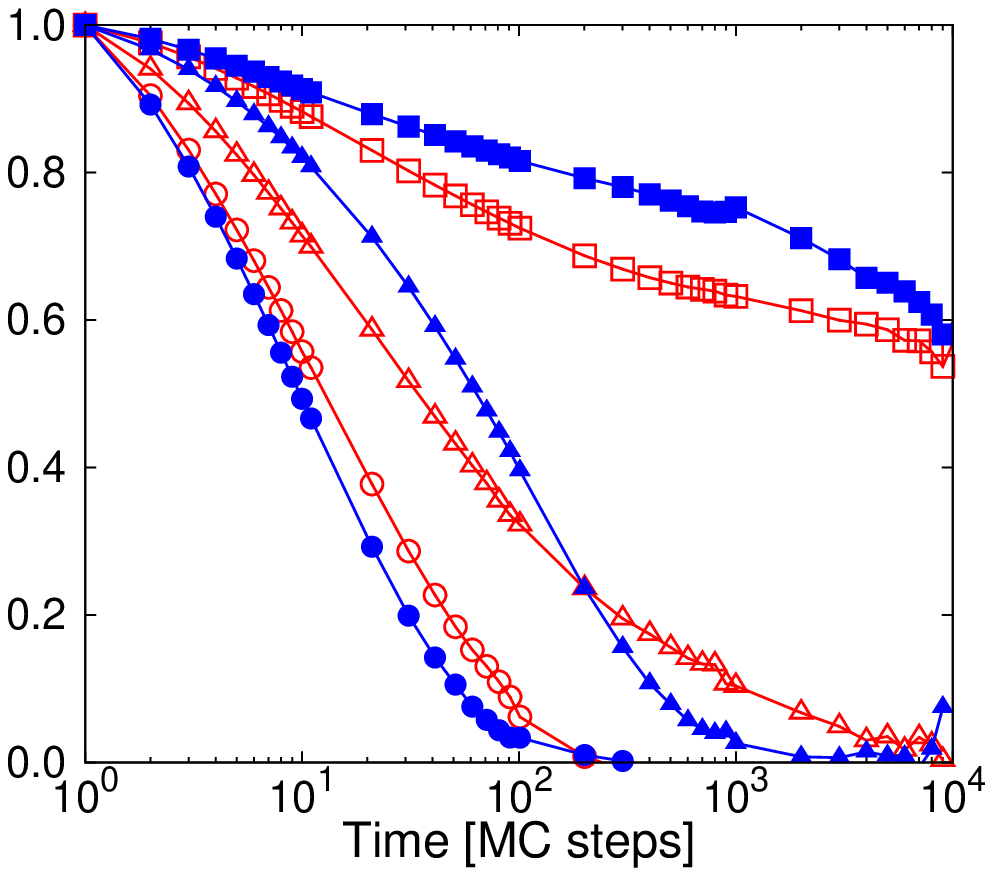}}
\caption{Autocorrelation functions of the average hydrogen bond index
  $M$ (a),  
and of the displacement of a molecule with respect to the center of its cell 
(b). The functions are calculated at the state points indicated in the
legend (with $T$ expressed in K and $P$ in GPa).  Pressures are as in
Fig.~\ref{RDF} and, for both pressures, 
the water monolayer is liquid for $T> 218$~K and solid for $T\leq
218$~K: LDC at $P=0.06$~GPa and hexatic at $P=0.24$~GPa.
}
\label{corr}
\end{figure}

For each quantity we define the correlation time $\tau$  as the time
at which the normalized correlation function Eq.~(\ref{Cm}) and
(\ref{Cr}) decay to $1/e$. Our calculations show that
at low $P$, the hydrogen bond correlation function $C_M$ (Fig.\ref{corr}a) is
exponential in the liquid phase, but has a 
non-exponential behavior in the LDC phase, with a correlation time
$\tau$ that largely increases for decreasing $T$, as expected in the
solid phase characterized by a well developed hydrogen bond network \cite{DynamicallySlow}. 
By increaing $P$, the number of hydrogen bonds largely decreases and
the hydrogen bond correlation function $C_M$ shows a much faster decay
to zero. Nevertheless, at high $P$ and low $T$ the function is
not-exponential consistent with the approach of a frozen state. 

For the translational autocorrelation function $C_r$ (Fig.\ref{corr}b) 
we find a much slower decay to zero at all the state points. For the
state points corresponding to the solid phases, at low
$T$ and any $P$, the function has an evident non-exponential behavior and an
extremely long correlation time $\tau$, two orders of magnitude
greater than $C_M$. This is consistent with the
arrested translational dynamics of the solid phases.



\begin{figure}
\centering
\subfloat[HB orientation variance]{\includegraphics[scale=0.45]{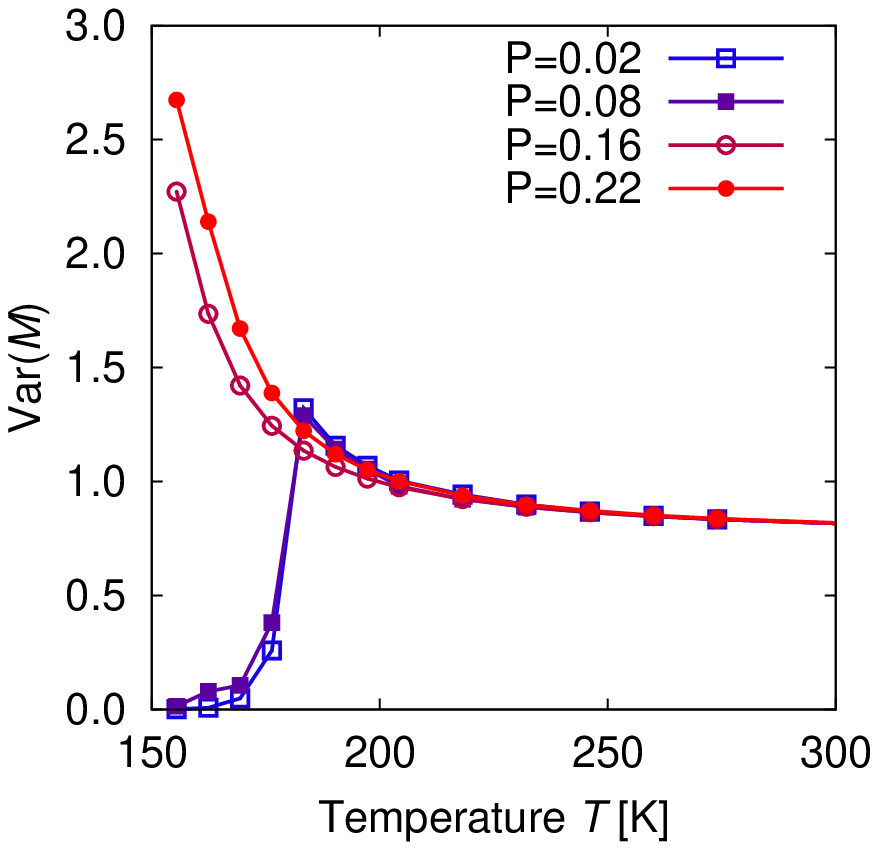}} 
\subfloat[Translational variance]{\includegraphics[scale=0.45]{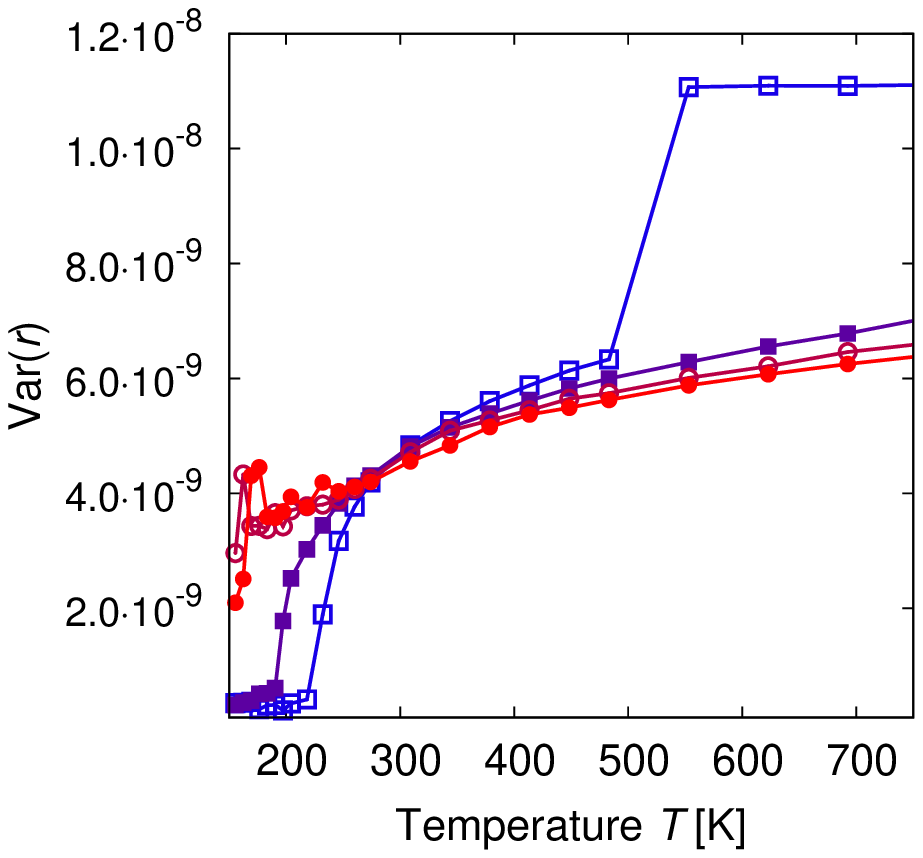}}
\caption{The variance of the average hydrogen bond index $M_i$ at (a), and 
of  the displacement $\delta \vec{r}_i$ of the molecule inside the cell
  at (b). At low $P$ both 
  display a discontinuous decrease at the liquid-LDC first-order phase 
  transition. At high $P$, around the liquid-hexatic second-order phase  transition,
$\langle \delta^2 M_i \rangle$ increases continuously, while 
$\langle \delta^2 (\delta \vec{r}_i) \rangle$ has a discontinuity.
At high $T$, $\langle \delta^2 (\delta \vec{r}_i) \rangle$ shows the
discontinuity associated to the liquid-gas first-order phase 
  transition. Pressures are expressed in GPa.}
\label{var}
\end{figure}

Next, we analyze the 
 behavior of the variance 
$\langle \delta^2 M_i \rangle$ of $M_i$,
 and $\langle \delta^2 (\delta \vec{r}_i) \rangle$ of
$\delta \vec{r}_i$, defined as
 the normalization factors of  Eq.s~(\ref{Cm}) and (\ref{Cr}),
 respectively (Fig.\ref{var}).

As expected,   at low $P$  both variances have discontinuties at the
temperature of the liquid-LDC first-order phase  transition. 
The increase of $\langle \delta^2 M_i \rangle$ for decreasing
$T$ is the consequence of  the increase of the fluctuations in the
hydrogen bonds network at low $T$. 

On the other hand, the
translational variance decreases for decreasing $T$ and presents
discontinuities at the first-order phase transitons, e. g. being
crystal at 
$T<274$ K, gas at 
$T>483$ K and a liquid at the intermediate values
at $P=0.02$~GPa.
At high $P$, instead, the slowing down of the translational dynamics
occurring at the liquid-hexatic coexistence is marked by a non
monotonic  $\langle \delta^2 (\delta \vec{r}_i) \rangle$,
suggesting an out of of equilibrium behavior at very low $T$.

\section{Conclusions}

We study by efficient Monte Carlo simulations a coarse-grained model
for a water monolayer in hydrophobic nanoconfinement 
and find two forms of ice at low $T$. At low pressure,  the model
reproduces the occurrence of low-density-crystal (LDC) ice with hydrogen bonds forming a
square network, as observed with detailed molecular dynamics
simulations. At high pressure, where detailed molecular dynamics
simulations are not available, we find a hexatic ice, separated from the
liquid phase by a second-order phase transition. Our structural
analysis shows that the hexatic phase has solid-like long-range orientational
order and liquid-like short-range translational order.

By studing the autocorrelation functions and the variances of bonding
and translational parameters, we observe different behaviors at low
and high $P$ that we can relate to the thermodynamic phases. 
The dynamics at low $T$ become slow at any pressure, and possibly
fall out of equilibrium at low $T$ in the hexatic phase, before
entering the high-density-crystal (HDC) ice phase.

\section*{Aknowledgement}
We thank the spanish MICINN grant FIS2009-10210 (Co-financed FEDER).


\begin{thebibliography}{99}

\bibitem{EfficientMC}
Djamal Bouzida, Shankar Kumar, and Robert~H. Swendsen.
\newblock Efficient monte carlo methods for the computer simulation of
  biological molecules.
\newblock {\em Phys. Rev. A}, 45(12):8894--8901, Jun 1992.

\bibitem{Hexatic}
Kun Chen, Theodore Kaplan, and Mark Mostoller.
\newblock Melting in two-dimensional lennard-jones systems: Observation of a
  metastable hexatic phase.
\newblock {\em Phys. Rev. Lett.}, 74(20):4019--4022, May 1995.

\bibitem{CooperativeDomains}
Jeffrey~R. Errington, Pablo~G. Debenedetti, and Salvatore Torquato.
\newblock Cooperative origin of low-density domains in liquid water.
\newblock {\em Phys. Rev. Lett.}, 89(21):215503, Oct 2002.

\bibitem{DynamicallySlow}
G.~Franzese and F.~de~los Santos.
\newblock Dynamically slow processes in supercooled water confined between
  hydrophobic plates.
\newblock {\em J. Phys.: Condens. Matter}, 21:504107, 2009.

\bibitem{TheoryDensity}
G.~{Franzese} and H.~{Eugene Stanley}.
\newblock {A theory for discriminating the mechanism responsible for the water
  density anomaly}.
\newblock {\em Physica A Statistical Mechanics and its Applications},
  314:508--513, November 2002.

\bibitem{IntramolecularCoupling}
Giancarlo Franzese, Manuel~I. Marqu\'es, and H.~Eugene~Stanley.
\newblock Intramolecular coupling as a mechanism for a liquid-liquid phase
  transition.
\newblock {\em Phys. Rev. E}, 67(1):011103, Jan 2003.

\bibitem{FoodBio}
V.~Bianco G.~Franzese and S.~Iskrov.
\newblock Water at interface with proteins.
\newblock {\em Food Biophysics}, Dec 2010.

\bibitem{Thermodynamics}
Pradeep Kumar, Sergey~V. Buldyrev, Francis~W. Starr, Nicolas Giovambattista,
  and H.~Eugene Stanley.
\newblock Thermodynamics, structure, and dynamics of water confined between
  hydrophobic plates.
\newblock {\em Phys. Rev. E}, 72(5):051503, Nov 2005.

\bibitem{TwoCrossovers}
M.~G. {Mazza}, K.~{Stokely}, S.~E. {Pagnotta}, F.~{Bruni}, H.~E. {Stanley}, and
  G.~{Franzese}.
\newblock {Two dynamic crossovers in protein hydration water and their
  thermodynamic interpretation}.
\newblock {\em ArXiv e-prints}, July 2009.

\bibitem{NpTMC}
I.~R. McDonald.
\newblock Npt-ensemble monte carlo calculations for binary liquid mixtures.
\newblock {\em Mol. Phys.}, 23(1):41--58, 1972.

\bibitem{MeltingLJ}
Alexander~Z. Patashinski, Rafal Orlik, Antoni~C. Mitus, Bartosz~A. Grzybowski,
  and Mark~A. Ratner.
\newblock Melting in 2d lennard-jones systems: What type of phase
  transition?†.
\newblock {\em The Journal of Physical Chemistry C}, 114(48):20749--20755,
  2010.

\bibitem{IceXV}
Christoph~G. Salzmann, Paolo~G. Radaelli, Erwin Mayer, and John~L. Finney.
\newblock Ice xv: A new thermodynamically stable phase of ice.
\newblock {\em Phys. Rev. Lett.}, 103(10):105701, Sep 2009.

\bibitem{Soper}
A.~K. Soper.
\newblock Structural transformations in amorphous ice and supercooled water and
  their relevance to the phase diagram of water.
\newblock {\em Mol. Phys.}, 106:2053--2076, 2008.

\bibitem{WaterStructure}
Alan~K. Soper and Maria~Antonietta Ricci.
\newblock Structures of high-density and low-density water.
\newblock {\em Phys. Rev. Lett.}, 84(13):2881--2884, Mar 2000.

\bibitem{HBCoop}
K.~{Stokely}, M.~G. {Mazza}, H.~E. {Stanley}, and G.~{Franzese}.
\newblock Effect of hydrogen bond cooperativity on the behavior of water.
\newblock {\em Proceedings of the National Academy of Science}, 107:1301--1306,
  January 2010.

\bibitem{OptimumMC}
J~Talbot, G~Tarjus, and P~Viot.
\newblock Optimum monte carlo simulations: some exact results.
\newblock {\em Journal of Physics A: Mathematical and General}, 36(34):9009,
  2003.

\bibitem{Freezing}
Thomas~M. Truskett, Salvatore Torquato, Srikanth Sastry, Pablo~G. Debenedetti,
  and Frank~H. Stillinger.
\newblock Structural precursor to freezing in the hard-disk and hard-sphere
  systems.
\newblock {\em Phys. Rev. E}, 58(3):3083--3088, Sep 1998.

\bibitem{BilayerIce}
R.~{Zangi} and A.~E. {Mark}.
\newblock Bilayer ice and alternate liquid phases of confined water.
\newblock {\em Journal of Chemical Physics}, 119:1694--1700, July 2003.

\bibitem{Monolayer}
Ronen Zangi and Alan~E. Mark.
\newblock Monolayer ice.
\newblock {\em Phys. Rev. Lett.}, 91(2):025502, Jul 2003.

\end{thebibliography}
\end{document}